\documentclass[aps,twocolumn,prl,floatfix,superscriptaddress,showpacs]{revtex4-1}
\usepackage[utf8]{inputenc} 

\usepackage{graphicx}
\usepackage{dcolumn}
\usepackage{bm}

\usepackage{braket}
\usepackage{amsfonts}
\usepackage{amsmath}
\usepackage{amssymb}

\usepackage{color,soul}

\usepackage{wasysym}
\usepackage{mathrsfs}
\usepackage{times}

\usepackage[dvipsnames,svgnames,table]{xcolor}

\usepackage{hyperref}
\usepackage{fancyhdr}
\usepackage{float}

\usepackage[english]{babel}

\hypersetup{
pdfstartview={FitH},% fits the width of the page to the window
colorlinks=true,    % false: boxed links; true: colored links
linkcolor=black,    % color of internal links
citecolor= black,   % color of links to bibliography
filecolor=black,    % color of file links
urlcolor=black      % color of external links
}

\newcommand{\bea}{\begin{eqnarray}}
\newcommand{\eea}{\end{eqnarray}}

\usepackage{lipsum}

\begin{document}

%%---------------------------------
%% Encabezado
%%---------------------------------
\title{Spin precession and laser-induced spin-polarized photocurrents}

\author{Esteban A. Rodríguez-Mena}
\affiliation{Univ. Grenoble Alpes, CEA, IRIG-MEM-L\_Sim, Grenoble, France.}

\author{Matías Berdakin}
\email{matiasberdakin@unc.edu.ar}
\affiliation{Consejo Nacional de Investigaciones Científicas y Técnicas (CONICET), Instituto de Investigaciones en Fisicoquímica de Córdoba (INFIQC), X5000HUA, Córdoba, Argentina}
\affiliation{Universidad Nacional de Córdoba, Facultad de Ciencias Químicas, Departamento de Química Teórica y Computacional, X5000HUA, Córdoba, Argentina}
\affiliation{Universidad Nacional de Córdoba, Centro Láser de Ciencias Moleculares, X5000HUA, Córdoba, Argentina}

\author{Luis E. F. Foa Torres}%∗^*
\email{luis.foatorres@uchile.cl} 
\affiliation{Departamento de F\'{\i}sica, Facultad de Ciencias F\'{\i}sicas y Matem\'aticas, Universidad de Chile, Santiago, Chile}

%%---------------------------------
%% Abstract
%%---------------------------------

\begin{abstract}
\section{abstract}

Controlling spin currents in topological insulators (TIs) is crucial for spintronics but challenged by the robustness of their chiral edge states, which impedes the spin manipulation required for devices like spin-field effect transistors (SFETs). We theoretically demonstrate that this challenge can be overcome by synergistically applying circularly polarized light and gate-tunable Rashba spin-orbit coupling (rSOC) to a 2D TI. Laser irradiation provides access to Floquet sidebands where rSOC induces controllable spin precession, leading to the generation of one-way, switchable spin-polarized photocurrents—an effect forbidden in equilibrium TIs. This mechanism effectively realizes SFET functionality within a driven TI, specifically operating within a distinct Floquet replica, offering a new paradigm for light-based control in topological spintronics.

\begin{description}
\item[Keywords]
Spin photocurrents, Topological insulators , two-dimensional materials, spin-orbit coupling   
\end{description}

\end{abstract}

\date{\today}
\maketitle

%%---------------------------------
%% Cuerpo
%%---------------------------------

\textit{Introduction.--} The quest for switchable spin-polarized currents, a cornerstone of spintronics, gained significant momentum with Datta and Das's groundbreaking 1980s proposal \cite{Datta_Das} for an electronic analog of the electro-optic modulator, later termed the spin-field effect transistor (SFET)~\cite{datta2018we}. Ahead of its time with respect to Rashba spin-orbit coupling (rSOC) manipulation and 2D materials, their concept envisioned the use of gate-tunable rSOC to control spin precession in a channel between ferromagnetic contacts, thereby modulating the spin polarization of the output current. The experimental realization two decades later \cite{koo2009control} confirmed the principle's validity but underscored the inherent challenges.

Separately, the discovery of topological insulators (TIs) \cite{Bernavig2006} introduced materials with intrinsically robust and chiral edge spin channels~\cite{Hasan2010}, seemingly ideal platforms for dissipationless spin-based devices~\cite{ortmann2015topological}. Yet, this very robustness poses a significant hurdle: manipulating transport---specifically, achieving the controllable spin-flipping required for SFET-like action---is fundamentally difficult in TIs due to the constraints of spin-momentum locking. While progress like the topological field-effect transistor~\cite{collins_electric-field-tuned_2018} addresses charge current control, realizing SFET functionality within a TI remains an outstanding challenge. Implementing an SFET directly appears to be a Herculean task, as flipping spin while maintaining propagation direction would typically require switching the physical edge channel itself.

\begin{figure}[htp]
    \centering
    \includegraphics[width=0.95\linewidth]{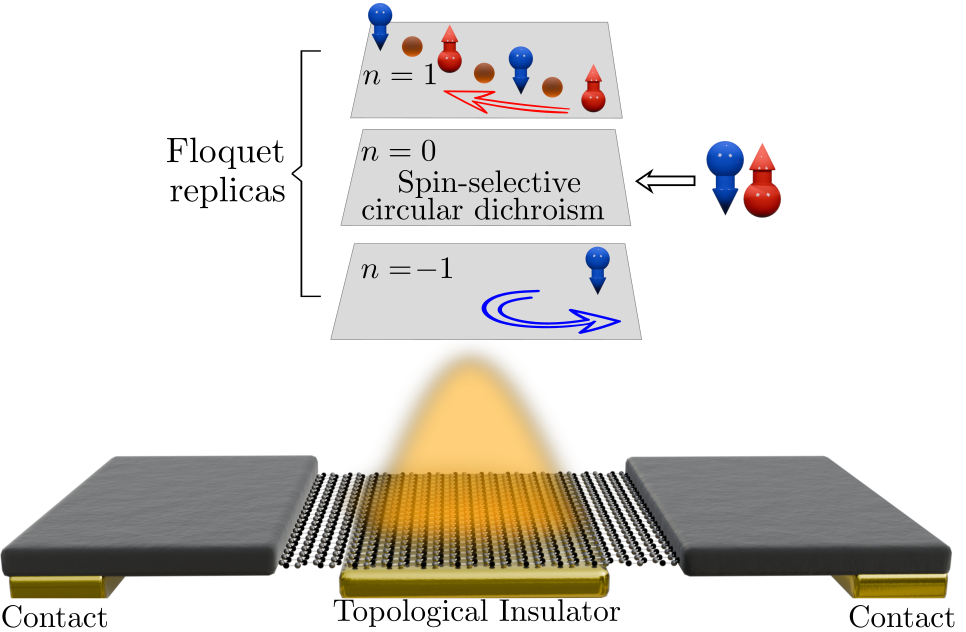}
    \caption{Device scheme. A 2D TI  is connected to two graphene leads. The central region is irradiated with circularly polarized light and an external gate allows tuning the rSOC. In the driven regime, the Hamiltonian unfolds into replicas of the equilibrium system, coupled by the laser field. These replicas represent the process of absorption and emission of $n$ multiples of $\hbar\Omega$. In 2D TIs, the spin-selective circular dichroism couples the spin components to different inelastic processes.}
    \label{fig:device_scheme}
\end{figure}

Strong laser fields offer a powerful toolkit for dynamically tuning material properties~\cite{noauthor_quantum_2020,basov_towards_2017}, capable of inducing phenomena ranging from Floquet-Bloch band gaps~\cite{lopez2008,syzranov2008,oka_photovoltaic_2009,kibis2010,calvo_app,mahmood_selective_2016,wang_observation_2013,setefan_floquet,Gedik2024floquet,liu2024}, engineering Floquet moiré patterns~\cite{Calvo_2025}, novel topological phases~\cite{oka_photovoltaic_2009, rodriguez-mena_topological_2019, rudner_band_2020,Perez-Piskunow_2014,sentef2015,nag2021,ke2024,mciver_light-induced_2020} to generating spin-polarized photocurrents in TIs via spin-selective light-matter interactions~\cite{mciver2012,berdakin_spin-polarized_2021,Ezawa2013}. However, harnessing light to achieve the specific \textit{controllable spin precession} needed for SFET-like switching within the topologically protected states demands a novel mechanism beyond these effects.

Here, we demonstrate that precisely combining circularly polarized laser irradiation with gate-controlled rSOC provides this mechanism. We show that this interplay enables the generation of one-way, switchable spin-polarized photocurrents, as illustrated schematically in Fig.~\ref{fig:device_scheme}. This occurs through controlled spin precession mediated by laser-induced hybridization with spin-depolarized Floquet sidebands, effectively realizing a light-driven SFET functionality within a specific Floquet replica of the topological insulator---a feat previously considered unattainable in these systems.

\begin{figure}[htp]
    \centering
    \includegraphics[width=1\linewidth]{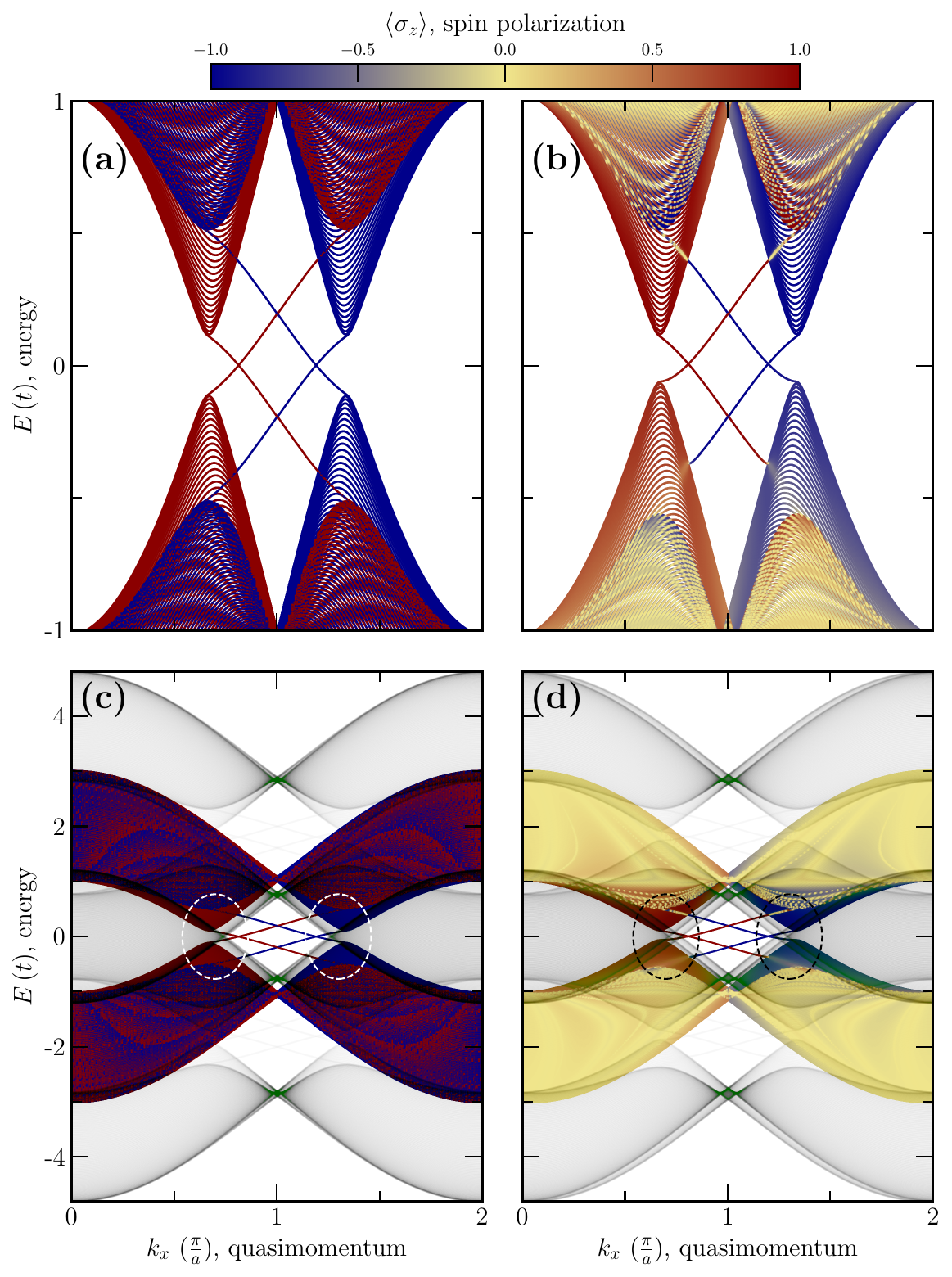}
    \caption{Band structure for Kane-Mele zig-zag ribbon as function of the rSOC $\lambda_{\mathrm{R}}=0,\,0.05$ for (a) and (b), respectively. (a) The spin polarization is completely defined, and behaving effectively as a two-fold copy of Haldane model. (b) As rSOC increase, edge states remain spin-polarized for a large range of amplitudes, but the bulk bands lose this feature. (c, d) Floquet spectrum in the decoupled limit ($\xi=0$) with $\hbar\Omega=1.5t$. The remaining parameters are set to $\mu_i=\pm 0.2$ and $\lambda_{\mathrm{SO}}=0.06$. Parameters are only set for illustration purposes. The white (black) dashed circles indicate possible new transitions between the spin-polarized edge states with the spin-polarized (depolarized) continuum of replicas once the coupling is enabled. The scale encodes the spin polarization degree $\langle \sigma_z \rangle \equiv \left|  \left< \uparrow|\psi(k_x)\right> \right|^2 - \left|  \left< \downarrow|\psi(k_x)\right> \right|^2 $ ranging from $-1$ to $+1$ for spin down and up, respectively.}
    \label{fig:km_rashba}
\end{figure}

\emph{Hamiltonian model for the irradiated Kane-Mele system.-{}-}
We start with a tight-binding description of TIs. For this, we employ the Kane and Mele Hamiltonian~\cite{kane_quantum_2005}, widely used as a general model to describe 2D TIs.~\cite{ezawa_monolayer_2015} It consists of a bipartite honeycomb lattice with staggered potential, intrinsic SOC interaction, and rSOC
\begin{align}
\mathcal{H}_{\text{KM}} & =\sum_{i}\mu_{i}\hat{c}_{i}^{\dagger}\hat{c}_{i}-t\sum_{\left<ij\right>}\hat{c}_{i}^{\dagger}\hat{c}_{j},\nonumber \\
 & +\mathrm{i}\lambda_{\mathrm{SO}}\sum_{\left<\left<ij\right>\right>}\nu_{ij}\hat{c}_{i}^{\dagger}\hat{s}_{z}\hat{c}_{j}+\mathrm{i}\lambda_{\mathrm{R}}\sum_{\left<ij\right>}\hat{c}_{i}^{\dagger}\left(\mathbf{s}\times\hat{\mathbf{d}}_{ij}\right)_{z}\hat{c}_{j}
\end{align}
\begin{figure*}
    \centering
    \includegraphics[width=0.8\linewidth]{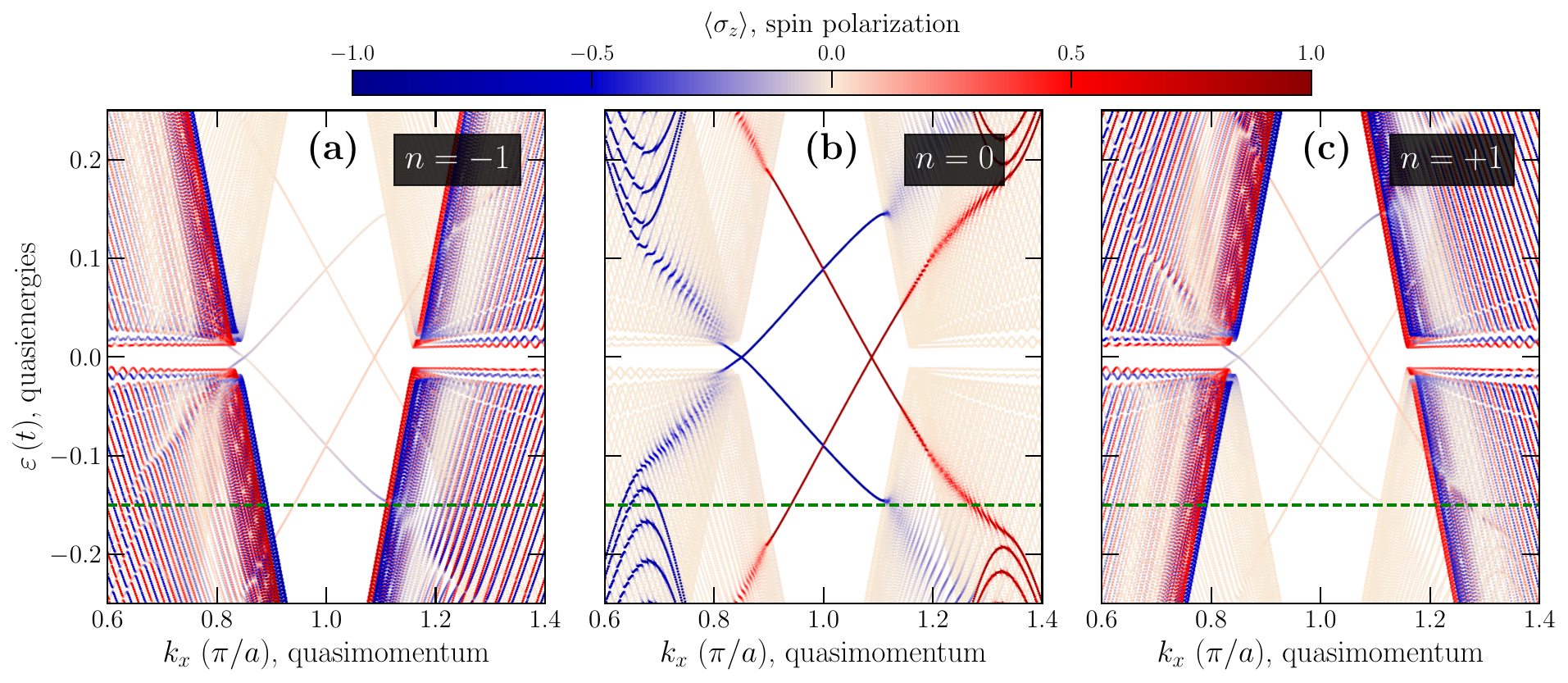}
    \caption{ Band structure for a circularly polarized laser irradiated Kane-Mele ribbon. Each panel presents the spectrum of the processes with $n=\pm 1, 0$. The color scale encodes the spin polarization over a specific Floquet replica, as indicated in each of the labels. The irradiation parameters are  $\hbar\Omega = 1.5$, $\phi=\pi/2$, and $\xi = 0.15$, while the remaining parameters remain fixed as described in the Hamiltonian section.}
    
    \label{fig:irrad_bands}
\end{figure*}

\noindent with $\hat{c}_{i}$ the fermionic creation operator at the site $\mathbf{r}_{i}$ and $\hat{s}_{z}$ the spin angular momentum operator for the electrons in $z$, the out-of-plane direction. Single and double angular brackets denote first and second nearest neighbors couplings. The quantities
$\mu_i$, $t$, $\lambda_{\mathrm{SO}}$ and $\lambda_{\mathrm{R}}$ stand for the staggered potential, nearest-neighbors coupling, amplitude of intrinsic SOC and rSOC, respectively. We set $t=1.6\,\mathrm{eV}$ as our unit of energy, and, except stated differently, the rest of the parameters take values compatible with Germanene ($\lambda_{\mathrm{SO}}=0.05$, $\lambda_{\mathrm{R}}=0.006$, and $\mu_{i}=\pm0.1$)~\cite{ezawa_monolayer_2015}.

The laser electric field enters the tight-binding Hamiltonian through Peierls' substitution~\cite{calvo_non-perturbative_2013,calvo_app}, modifying all inter-site couplings by a phase factor:

\begin{equation}
t_{ij}\hat{c}_{i}^{\dagger}\hat{c}_{j}\to t_{ij}\exp\left(\mathrm{i}\frac{2\pi}{\Phi_{0}}\int_{\mathrm{r}_{j}}^{\mathrm{r}_{i}}\mathbf{A}(t)\cdot d\mathbf{r}\right)\hat{c}_{i}^{\dagger}\hat{c}_{j},\label{eq:peierls}
\end{equation}
where $\mathbf{A}(t)$ is the vector potential, $t_{ij}$ the bare hopping amplitude, and
$\Phi_{0}$ is the quantum magnetic flux. Since a spin-selective optical response can be achieved only by irradiation with circularly polarized light (i.e. through the so-called spin-selective circular dichroism effect \cite{ghalamkari_perfect_2018}), our study focuses on the action of circularly polarized light perpendicular to the two-dimensional material. We consider: $\mathbf{A}(t)=A_{0}\cos\left(\Omega t\right)\hat{x}+A_{0}\cos\left(\Omega t+\phi\right)\hat{y}$,
with $\phi=\pm\pi/2$, the laser strength characterized by the dimensionless parameter $\xi\equiv2\pi A_0a/\Phi_0$, being $a=4.02\,\mathrm{\r{A}}$ the lattice constant, set as our unit of length. Since the resulting Hamiltonian is time-periodic, the problem can be tackled using the Floquet formalism as described in the following section.

\emph{Floquet theory.-- } After incorporating the Peierls' phase given
by (\ref{eq:peierls}), the resulting Hamiltonian is $T$-periodic $\mathcal{H}(t)=\mathcal{H}(t+T)$.
The Floquet theorem can be used to obtain the spectrum and stationary transport properties, including scattering wave functions of an irradiated device. The Floquet theorem guarantees a complete set of eigenstates of the form:

\begin{equation}
    \psi_{s, \alpha}(\mathbf{r},t)=\exp(-\mathrm{i}\varepsilon_\alpha t/\hbar)\phi_{s, \alpha}(\mathbf{r},t),
\end{equation}
\noindent where $\mathbf{r}$ are the orbital degrees of freedom, $s$ the spin degree of freedom, and $\varepsilon_{\alpha}$ are the quasienergies, defined within the first Floquet zone $-\hbar \Omega/2 \leq \varepsilon_{\alpha} \leq \hbar\Omega/2$. The $\phi_{s, \alpha}$ are the time-periodic Floquet states satisfying $\mathcal{H}_F\phi_\alpha(\mathbf{r},s,t)=\varepsilon_{\alpha}\phi_{\alpha}(\mathbf{r},s,t)$ where $\mathcal{H}_F\equiv \mathcal{H}-\mathrm{i}\hbar \partial_t$ is the Floquet Hamiltonian. The matrix elements of $\mathcal{H}_F$ are given by:

\begin{align}
\Braket{m,\lambda|\mathcal{H}_{F}|n,\sigma} & =\frac{1}{T}\int_{0}^{T}\Braket{\lambda|\mathcal{H}(\mathbf{r},t)|\sigma}e^{-\mathrm{i}(m-n)\Omega t}\nonumber \\
 & +m\hbar\Omega\delta_{nm}\delta_{\lambda\sigma}\label{eq:mat-elemets-hf},
\end{align}
\noindent with the latin indices reserved for the Floquet replicas and the greek indices for the Hilbert space that describes the Hamiltonian in the static limit. Thus, the resulting eigenvalue problem is represented in the direct product space (usually named Floquet or Sambe space ~\cite{sambe_steady_1973}) $\cal{R} \otimes \cal{T}$ where $\cal{R}$ is the usual Hilbert space and $\cal{T}$ the space of T-periodic functions spanned by $\langle t |n \rangle = \exp(\mathrm{i}n\Omega t)$, where $n$ is the ``replica" index associated with the number of photons. We employ as many Floquet replicas as needed to ensure the convergence of any physical observable. 

Here we consider a two-terminal device (labeled $\alpha$ and $\beta$), where the scattering region is an irradiated TI, modeled by a Kane-Mele Hamiltonian in the topological phase. The leads are made of graphene (see Fig.~\ref{fig:device_scheme}). Within Floquet theory, an electron entering the scattering zone through lead $\alpha$ in the so-called reference replica ($n_0=0$) can exit through $\beta$ exchanging $n$ photons. Since only the scattering zone is irradiated, in the leads the replicas are decoupled, such that the total transmission probability from the lead $\alpha$ to $\beta$ at quasienergy $\varepsilon$, $\cal{T}_{\beta, \alpha}(\varepsilon)$, is obtained by summing the probabilities of each of the independent processes: 
\begin{align}
{\cal{T}}_{\beta,\alpha}(\varepsilon) = \sum_{n}  {\cal{T}}_{\beta,\alpha}^{(n)} (\varepsilon).
\end{align}
The spectral and transport properties were obtained through an implementation of Eq.~\ref{eq:mat-elemets-hf} based on the Kwant module \cite{Groth_2014}.

\begin{figure*}[t]
    \centering
    \includegraphics[width=0.9\linewidth]{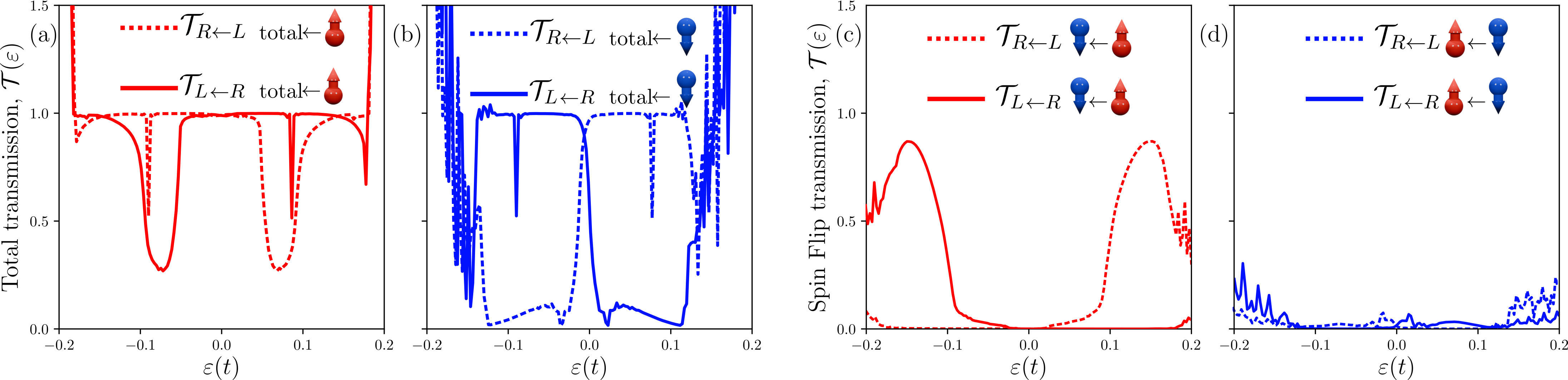}
    \caption{Total transmission probabilities (a, b) and spin-flip contributions (c, d) to the transport properties. Spin-up (spin-down) incidence is denoted by red (blue) lines, while the incidence direction is indicated by solid ($L\leftarrow R$) and dashed ($R\leftarrow L$) line styles. The device dimensions are  $500a \times 200a$ and the irradiation parameters are  $\hbar\Omega = 1.5$, $\phi=\pi/2$ and $\xi = 0.15$. The rest of the parameters remain the same as those described in the Hamiltonian section.}
    \label{fig:transport}
\end{figure*}
\emph{Hybridization with continuum of replicas.--} Rashba spin-orbit coupling (rSOC) allows spin-flip processes. Remarkably, its presence does not disrupt the spin polarization of the topologically protected edge states in the Kane-Mele Hamiltonian. Indeed, the topological states are robust against a moderately strong rSOC ($\lambda_{\mathrm{R}} < 2\sqrt{3}\lambda_{\mathrm{SO}}$)~\cite{kane_quantum_2005, kane_z_2005}.
In Fig. \ref{fig:km_rashba} the band structure of the Kane-Mele Hamiltonian is presented, using a color scale to highlight its spin-polarization. Panels (a, b) show the spin polarization of the spectrum with increasing rSOC, being $\lambda_{\mathrm{R}}=0$ for (a) and $\lambda_{\mathrm{R}}=0.05$ for (b). As can be seen, there is no significant difference among the spin-polarization of the edge channels under the two conditions. However, as $\lambda_{\mathrm{R}}$ increases, the spin polarization within the bulk bands fades away, making them spin unpolarized. Although the spin polarization of bulk bands remains unspecified—apparently rendering them unsuitable for spintronic applications—we will demonstrate that this very feature is crucial for enabling the SFET mechanism in irradiated TIs. 

The Floquet spectrum consists of replicas of the stationary Hamiltonian, shifted by integer multiples of $\hbar\Omega$. Panels (c) and (d) of Fig. \ref{fig:km_rashba}  (green curves) display these Floquet sidebands in the decoupled limit ($\xi=0$), corresponding to the rSOC conditions shown in panels (a) and (b), respectively. Dashed circles mark the regions where edge states overlap with Floquet replicas. Since these states belong to the bulk of the replicas, a finite rSOC ($\lambda_{\mathrm{R}}\neq0$) creates an unpolarized spin environment to interact with the spin channels.

In the non-decoupled limit, laser irradiation induces specific hybridization with the replicas, providing a knob to tune the band structure of the system on demand.~\cite{Foa_2014, giovannini2020floquet} Similar approaches have been employed in \cite{dal_lago_one-way_2017, berdakin_directional_2018} to harness hybridization, albeit with a different mechanism, where the continuum is supplied by a second physical layer. Additionally, hybridization with a continuum of states is known to allow the perturbation of the native topological chiral states present in TIs~\cite{berdakin_spin-polarized_2021}. In Fig. \ref{fig:irrad_bands} the band structure of a Kane-Mele ribbon irradiated with circularly polarized light is shown. For the sake of simplicity, we separate the spectra of the processes with $n=\pm 1,0$. Since we are interested in the effect of irradiation on the transport properties of the topological edge states, we focus on the energy range spanned by the topological gap of the unperturbed material. To highlight the basic ingredients needed for the light-induced SFET mechanism to occur in TIs, we focus our analysis on the edges states near $E_F\approx -0.15$ (green line). For spin-down (blue), at $n=0$ modes propagate with nearly zero group velocity, backscattering is forbidden on this replica, and there are no states to couple conserving $k_x$ at $n=1$ replica. Hybridization with $n=-1$ is possible (see the fading blue color in $n=0$), but as we are going to discuss later, transport at this replica is suppressed due to a selection rule involving spins, the handedness of the laser, and the nature of the process (absorption or emission); for details, see Ref. \cite{berdakin_spin-polarized_2021}. On the other hand, in this range of energies, spin-up edge state (red) can hybridize both $n=1$ and $n=-1$ replicas that are spin depolarized in the vicinity of energy and momentum.  As a result, once a photon has been absorbed, the spin polarization is not well-defined anymore. In the following paragraphs, we study the outcome of lifting the spin-polarization restriction on the transport properties of an irradiated device.

\emph{Transport properties.--} Let us now examine the transport properties of the two-terminal setup (Fig.~\ref{fig:device_scheme}), where only the central TI region is illuminated and the leads remain in equilibrium. Our goal is to understand how the interplay between rSOC and laser irradiation influences the spin polarization of the transmitted charge. To this end, we performed transport simulations using spin-polarized incident modes.

Figure~\ref{fig:transport} presents the calculated transmissions. The spin-up (spin-down) incident mode is represented by a red (blue) line, whereas the direction of incidence is indicated by solid and dashed line styles. Panels (a) and (b) show the total transmission probability across the device, summed over all outgoing modes and final spin states. In contrast, panels (c) and (d) show the spin-flip contribution to transport.

The total transmission profiles in panels (a,b) qualitatively resemble those previously reported in Ref.~\cite{berdakin_spin-polarized_2021} for a similar setup without input spin selection. In the absence of irradiation, one expects reciprocal, perfectly quantized transmission plateaus corresponding to the number of edge states within the gap. However, Fig.~\ref{fig:transport}(a,b) reveals a stark deviation from this equilibrium behavior under illumination. One-way spin-polarized photocurrents emerge, driven by the interplay between broken symmetries in the Hamiltonian, coupling to inelastic Floquet channels, and the spin-selective dichroism effect identified previously~\cite{berdakin_spin-polarized_2021,Ezawa_2012}.

The panels of Fig.~\ref{fig:transport}(c,d) allow us to analyze how the total transmissions are built up. It can be noted that the total transmission cannot be explained solely by the spin-conserving contributions. In fact, the spin-flip contribution is far from negligible across a wide range of energies that span the gap, and its efficiency can be remarkably high (up to $\sim$ 0.9). These results can be understood by considering two key ingredients:
\begin{enumerate}
    \item The spin-selective dichroism effect observed in irradiated TIs generates spin-polarized photocurrents. In other words, circularly polarized laser irradiation allows one spin channel (in this case, up) to pass through the device, while the other (down—see the transport quench in panel (b)) is selectively backscattered via an inelastic process.~\cite{berdakin_spin-polarized_2021}
    \item As described earlier, the inelastic process responsible for forward scattering occurs within a spin-depolarized continuum of states. Therefore, once a photon is absorbed, the unpolarized environment enables spin precession, allowing spin-flips to occur.
\end{enumerate}
These are the key ingredients required to build a laser-driven analogue of the SFET in topological insulators. In the following paragraphs, we explore spin precession and wavelength commensuration in the irradiated device.

\begin{figure}
    \centering
    \includegraphics[width=\linewidth]{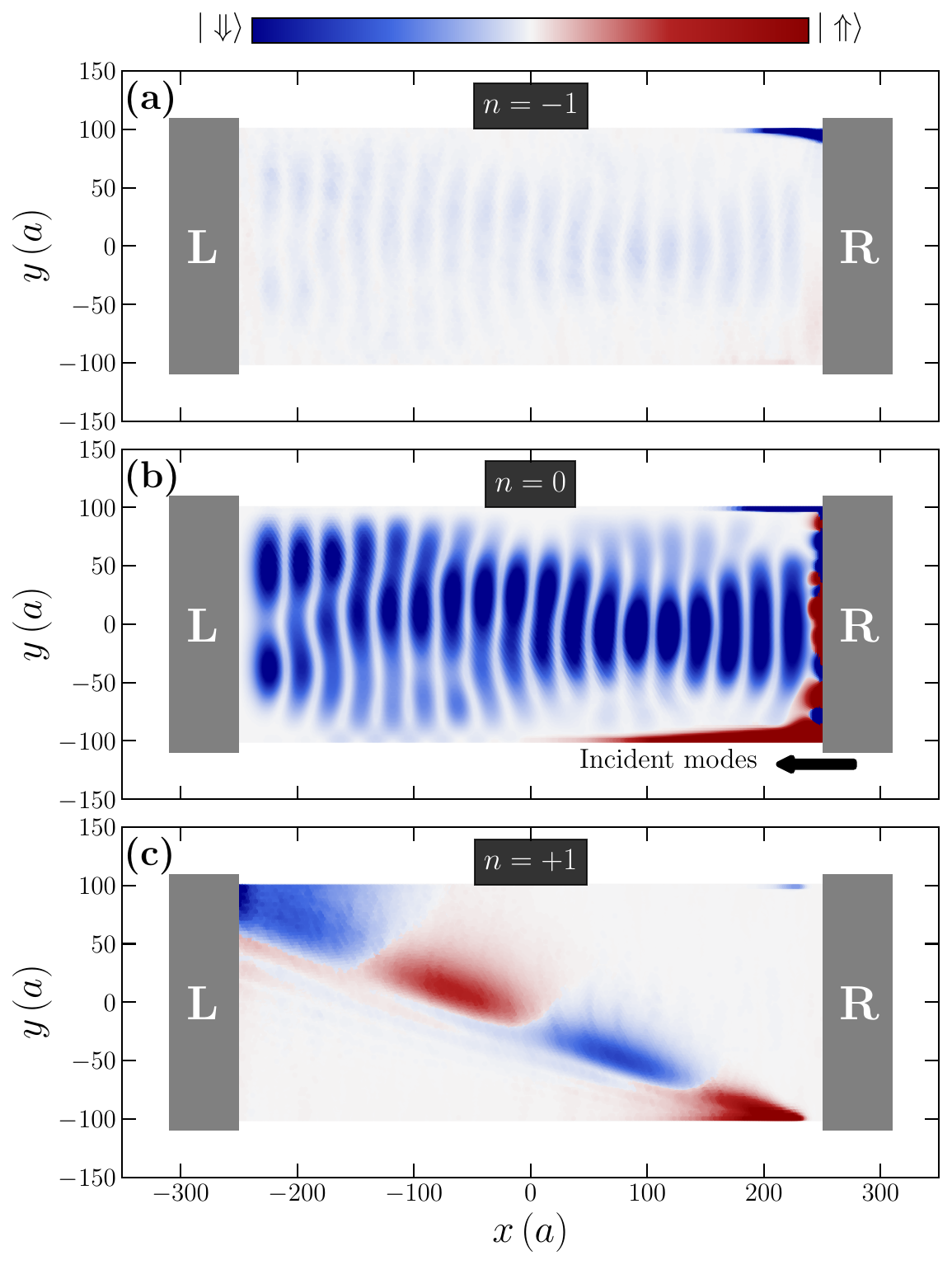}
    \caption{Scattering wave functions (SWF) a two-terminal configuration for a Fermi energy of $E_F\approx -0.15$. Color encodes the spin polarization and the index $n$ denotes the replica index. All the parameters are the same than in Fig.\ref{fig:transport}}
    \label{fig:setup}
\end{figure}

To gain a deeper insight into the spin precession within the irradiated device, we analyze the spatial distribution of the scattering wave function (SWF). Fig. \ref{fig:setup} shows the SWF obtained at $E_F\approx -0.15$. For clarity, the contribution of each replica to the SWF is plotted separately, with the spin polarization encoded in the color scale. These results further support the spin-flip mechanism described so far. In the elastic channel, spin-down transport is suppressed due to the cancellation of the edge state's group velocity as it reaches the bulk, while the spin-up contribution remains localized at the edge of the device until it is fully hybridized by the interaction with other replicas. The spin-selective dichroism effect implies that each spin is selectively hybridized with different inelastic processes \cite{berdakin_spin-polarized_2021}; in this case, spin-down and spin-up electrons couple with $n=-1$ and $n=1$ replicas, respectively. Within the $n=-1$ replica, spin-down transport is suppressed leaving backscattering as the only available channel. Conversely, in the $n=1$ replica, spin-up polarized electrons undergo Rabi oscillations, as clearly seen in Fig. \ref{fig:setup}(c). Once again, these results highlight that light enables SFET functionality in TIs. In the following section, the role of rSOC in the magnitude of the precession wavelength will be assessed, providing a rule of thumb to tune a device from a spin-conserving to a spin-switching regimen.

\emph{Wavelength estimation.--} As shown in Fig.~\ref{fig:setup}(c), the spin part of the SWF exhibits a distinctive Rabi oscillation pattern. Our task in this section is to elucidate the dependence of the precession wavelength on the rSOC, thus providing a heuristic argument to design a device that will behave as a spin-preserving or a spin-flipper one. 

\begin{figure}[htp]
    \centering
    \includegraphics[width=1\linewidth]{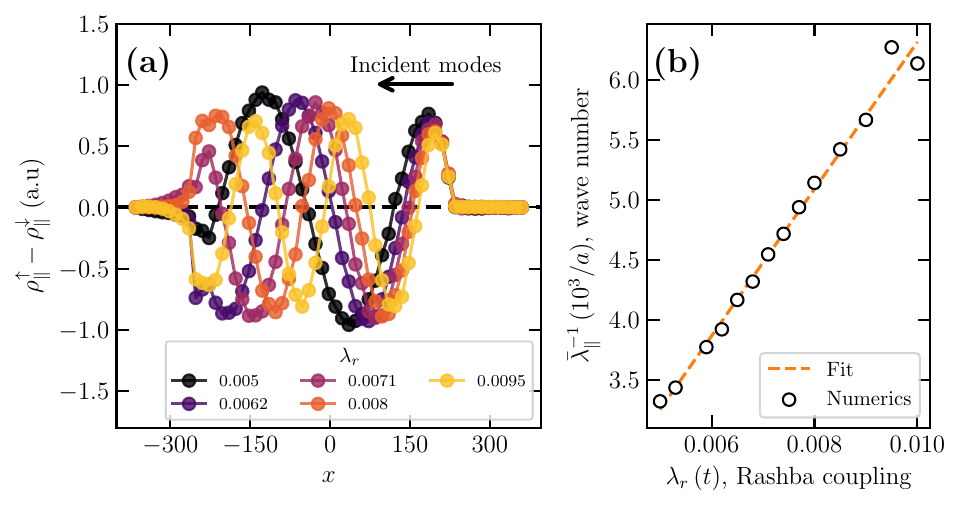}
    \caption{(a) Projection of the spin polarization of the SWF on the longitudinal axes of the device as a function of rSOC with $\rho_{\parallel}^{\uparrow, \downarrow}\equiv |\Psi^{\uparrow, \downarrow}_{\mathrm{SWF}}|^2$ the (up, down) wavefunction probability density. Oscillations can be readily observed exhibiting a Rabi-like oscillation pattering over the device within the replica $+1$.(b) Linear fitting of the oscillation wavelength as a function of the Rashba SOC. All the parameters are the same than in Fig.\ref{fig:transport}  
}
    \label{fig:linear_fitting}
\end{figure}

We measure the wavelength of the SWF using its parallel projection to the $x$ axis (see Fig. \ref{fig:linear_fitting}a). For a TI sample with fixed dimensions of $500a \times 200a$, we perform our simulations with varying rSOC amplitude. The wavenumber's dependence on the rSOC intensity exhibits a linear law $\lambda^{-1}_{\parallel}=1.57\lambda_{\mathrm{R}}$, as depicted in Fig. \ref{fig:linear_fitting}b. Since this occurs over the inelastic channel $n=+1$, it demonstrates that the external field favors the spin-flip processes by providing a continuum background allowing hybridization that does not preserve $z$-direction spin angular momentum of the edge states. The linear dependence on the wavelength with rSOC is a distinctive feature of the SFET, which agrees with the linear dependence of the spin phase shift proposed by Datta and Das in \cite{Datta_Das}. These results support the conclusion that light irradiation enables the SFET mechanism in TIs, which, by design do not support such spin-flipping processes.

We can establish a simple rule of thumb to determine the desired functionality of the device. Consider a sample of length
$L$. Since the wavelength spans a complete Rabi cycle, we define the ratio $n \equiv L/\lambda_{\parallel}$ It follows that if 
$n$ is an integer, the device will preserve the incident spin, whereas if $n$ is a half-integer, the device will act as a spin flipper. Given the linear dependence of the precession wavelength on rSOC, this rule provides a straightforward way to tune the device behavior by adjusting an external field to set the rSOC strength.

\textit{Final remarks.--}
Exploiting the unique properties of topological insulators (TIs) for advanced spintronic devices, such as the spin-field effect transistor (SFET), holds immense promise. However, realizing this potential is fundamentally challenged by the inherent robustness and spin-momentum locking of TI edge states, which typically prevents the direct spin manipulation required for SFET functionality. Driving materials out of equilibrium offers pathways to bypass such constraints and unlock new functionalities.

In this work, we have theoretically demonstrated a non-equilibrium strategy to overcome this limitation in 2D TIs by synergistically combining circularly polarized laser irradiation with gate-tunable Rashba spin-orbit coupling (rSOC). We established that the laser field provides access to engineered Floquet sidebands, creating an effective environment where rSOC induces controlled spin precession of the topological states. This process relies on the interplay between the spin-selective hybridization with these Floquet states (a dichroism effect) and the tunable spin-flipping capability of rSOC within this non-equilibrium manifold.

The key outcome is the generation of switchable, one-way spin-polarized photocurrents, which effectively replicates the operational principle of an SFET, but now dynamically realized within a specific Floquet replica. Our findings introduce a novel paradigm for achieving dynamic spin control in TIs by leveraging Floquet engineering to circumvent equilibrium restrictions. This light-induced SFET mechanism not only demonstrates the profound extent to which light can tailor quantum material properties but also opens avenues for designing optically-controlled topological spintronic components and further exploring non-equilibrium spin physics.

\begin{acknowledgments}
\section{acknowledgments}
We thank the support of FondeCyT (Chile) under grant number 1250751, and by the EU Horizon 2020 research and innovation program under the Marie-Sklodowska-Curie Grant Agreement No. 873028 (HYDROTRONICS Project). L. E. F. F. T. also acknowledges support from the ICTP through the Associates Programme and from the Simons Foundation through grant number 284558FY19. M. B. acknowledges financial support by Consejo Nacional de Investigaciones Cient\'ificas y T\'ecnicas (CONICET) under grant number PIBA 28720210100973CO, Agencia Nacional de Promoción Científica y Tecnológica under grant number 01-PICT 2022-2022-02-00277, and Secretar\'ia de Ciencia y Tecnolog\'ia de la Universidad Nacional de C\'ordoba (SECYT-UNC) under grant number 33820230100101CB. 

\end{acknowledgments}

\section{References}
\bibliographystyle{apsrev4-1_title}
\bibliography{referencias.bib}
\newpage

% \begin{figure}
%     \centering
%     \includegraphics[width=0.95\linewidth]{fig_toc.png}
%     \caption{TOC}
%     \label{fig:toc}
% \end{figure}

% \appendix*

\end{document}